# Open at the Top; Open at the Bottom; and Continually (but Slowly) Evolving

Russ Abbott
Department of Computer Science
California State University, Los Angeles
and The Aerospace Corporation
Los Angeles, Ca, USA
Russ.Abbott@GMail.com

**Abstract –** *Systems of systems differ from traditional systems in that they are open at the top, open at the bottom, and continually (but slowly) evolving. "Open at the top" means that there is no pre-defined top level application. New applications may be created at any time. "Open at the bottom" means that the system primitives are defined functionally rather than concretely. This allows the implementation of these primitives to be modified as technology changes. "Continually (but slowly) evolving" means that the system's functionality is stable enough to be useful but is understood to be subject to modification. Systems with these properties tend to be environments within which other systems operate—and hence are systems of systems. It is also important to understand the larger environment within which a system of systems exists.*

## 1 Introduction

The term *system of systems* has been in widespread use for at least a decade. As long ago as 1996 Maier [1] offered "Architecting Principles for Systems-of-Systems." In last year's IEEE Conference on Systems of Systems Engineering (October 2005) Jamshidi [2] listed six other definitions of system of systems, which he asserted were among the most common. Three date from 1998 or earlier.

Even though the term *system of systems* has been around for quite a while, we still seem to be struggling with the concept. Jamshidi quoted approvingly from the claim in Sage and Cuppan [3] that "there is no universally accepted definition of systems of systems." Most definitions of *system of systems* are not very helpful. Some are harmful.

An example of a not very helpful definition is Kotov's [4] (from 1997), which Jamshidi describes as his "favorite." It reads, "Systems of systems are large scale concurrent and distributed systems that are comprised of complex systems."

This definition does not offer much insight into either what a system of systems is or what distinguishes a *system of systems* from, say, a *collection* of (large scale concurrent and distributed) systems. The definition is flawed also in that it attempts to distinguish systems of systems from just plain systems—which themselves may be composed of subsidiary systems—on the grounds that the component systems that make up a system of systems must be complex. Not only is it unclear how a complex system is defined for the purpose of this definition, it is also unclear how this definition distinguishes a system of system from a plain old system whose (non-system) subsystems are complex.

The Defense Acquisition University's Defense Acquisition Guidebook [5] includes a discussion of system of systems engineering (section 4.2.6).

> System of systems engineering deals with planning, analyzing, organizing, and integrating the capabilities of a mix of existing and new systems into a system of systems capability greater than the sum of the capabilities of the constituent parts.

What is unfortunate about this definition is its focus on the capability provided by the system of systems as a functional entity. In doing so, it treats a system of systems as similar to any other system—defined in terms of a set of capabilities. According to this definition, the primary difference between a system of systems and a regular system is apparently that the systems of systems has systems (rather than subsystems?) as components. From this perspective, systems of systems are not qualitatively different from plain old systems. If there are no significant qualitative or structural differences between traditional systems and systems of systems, what really is all the fuss about?

In this paper we argue that properly understood a system of systems is qualitatively and structurally different from a traditional system and that the term *system of systems* should be reserved for this new perspective. As we elaborate below, a system of systems is best viewed not as a hierarchy built of component systems but as an environment within which other systems operate and which can support the addition of new systems that build on systems already in the environment. Furthermore to fully understand a system of systems not only must it be viewed as an environment for other systems, it must itself be understood in terms of the larger environment within which it and its participating systems exist. In other words, a system of systems perspective requires one to look outward from a system rather than inwards towards the system's hierarchical components.

## 2   Systems of systems are different

Our objective here is to describe how systems of systems are qualitatively and structurally different from traditional system and are not just a larger version of the same old hierarchical structure. We claim that a system of systems—when properly understood as an environment along with the systems operating within it—is open at the top, open at the bottom, and continually evolving—but slowly enough to be stable.

1. *Open at the top.* A system of systems is not defined in terms of some fixed top-level application. This distinguishes it from most systems as traditionally understood. Instead, a system of system enables the continual introduction of new applications. For example, although often quite primitive, the so-called *mashups* combine information from two or more websites. Many combine Google Maps with some other source. (See [6], which publishes a matrix of mashups, updated daily.) Burke [7] documented such connections more generally three decades ago.

2. *Open at the bottom.* There is no fixed bottom level for a system of systems. The lowest level of a system of system may be changed out from under it at any time. As an example, consider any communication stack. (As we shall discuss below, communication systems are prototypical systems of systems.) Signal transport is typically the lowest layer. Yet consider how wireless networks are replacing wired connections. If one were committed to wires as a fixed concrete bottom layer, this would not be possible. It is frequently the case that changes at the lowest level reflect influences from the environment within which the system of system is itself embedded—in this case improved technology developed in the world outside the communication system.

3. *Continually evolving, but slowly.* A system of systems is never finished. It evolves continually as the environment within which it operates changes. Systems of systems evolve in at least three ways. (a) Technology changes: wireless replaces wires. (b) Usage changes. New features are added, and existing features are modified. A system of systems must be able to support new uses of existing capa-bilities as well as the addition of new capabilities built on top of existing capabilities. (c) Standards and interfaces change. Even though stable standards and interfaces are important, it must be possible to change them as needs change and as our understanding of what we are attempting to standardize improves. Consider virtually any standard. One will find that it is continually (although not too frequently) re-issued as the purposes for which it is used evolve and as we understand better what it was intended to do.

Systems with these properties do not lend themselves to easy hierarchical control. On the other hand, systems of this sort are not completely formless. Any system, to be useful, must be able to perform specific functions at particular times. Systems of systems achieve this goal in that at any given time (a) they include a collection of (relatively stable) participating systems and (b) they implement a (relatively stable) set of standards and interfaces. But neither the set of participating systems nor the standards and interfaces are fixed forever. They evolve—but slowly. Thus the best way to think of a system of systems is as a collection of participating systems along with a set of shared standards and interfaces all of which evolve slowly enough so that making an investment in them is likely to be worthwhile.

What do we mean by a standard that is continually evolving but evolving slowly enough so that making an investment in it is worthwhile? Think of a person. Each of us changes physically from day to day. The atoms that make up our body one day are not the same atoms that make up our body the next. Moreover, we change our structures as well. We deteriorate with age, and we improve with learning. We are not isomorphic to our ourselves of a certain number of years ago—although there is generally a significant resemblance. Although we change continually, we change slowly enough that we find it worthwhile to invest in ourselves. There is a pace of change that we as human beings (and as organizations composed of human beings) can tolerate. Systems of systems change continually—often in fundamental ways. But the rate at which a systems of systems changes is for the most part slow enough that we can adjust and adapt.

Most changes to a system of systems will be slow—although an aggregation of small changes sometimes leads to a phase change and to punctuated equilibrium effects. But fast or slow, there will be changes. It is wishful thinking (or foolish wishing) to imagine that a system (or a system of system) (a) can be built according to a specification that is so perfect and complete and (b) will have an architecture that is so abstract and general that the system's fundamental framework will be eternal. Other than nature itself, no system has a permanent set of rules. And nature, the ultimate system of systems, is at the same time both eternal and always evolving.

## 2.1 Communication systems: the USPS

In this section we consider wide area communication networks—such as the United States Postal Service (USPS), the telephone infrastructure, and the internet—as prototypical systems of systems. They all exhibit the properties of being open at the top, open at the bottom, and always evolving, albeit slowly. Of the three systems mentioned we will examine the United States Postal Service since it is both the most longstanding and the least examined—at least in this context.

No one has ever accused the USPS (the people who bring us "snail mail") of being a high tech system. It was originally created under authorization by the US constitution, and it has existed for as long as the country. A pre-constitutional postal service existed in colonial times. The constitution authorized Congress to continue providing postal services. (See [8].) The USPS has been remarkable in both its stability and its adaptability.

The primary function of the USPS is to deliver mail (envelopes and packages) to addressees in the United States. One may think of this as defining its user interface: give the USPS an item (an envelop or package) with a valid address (and sufficient postage), and that item will be delivered to the indicated address. (Strangely no explicit formal statement of USPS services was found anywhere on the USPS website: http://www.usps.com/. It may be that the fundamental mission of the USPS is defined in the same way that English common law is defined, by tradition.)

Conceptually, what the USPS offers is a remarkably simple and intuitive service. All one needs to know are (a) how to address an item (the semantics), (b) how to enter an addressed item into the USPS system (the mechanism for providing input to the system), and (c) how to retrieve items from the system (the mechanism for retrieving output). The rest of the system may be understood as a black box. Nothing need be said about how the USPS goes about getting items from where they are received to their destinations.

*Open at the bottom.* When we say that a system of systems is open at the bottom, we refer to the system's ability to change its lowest level of functionality. In the case of the USPS, the system was able to re-implement its internal mail sorting system to make use of technologies (optical character recognition devices, for example) that didn't exist when the system came into existence.

Another illustration of the way in which the USPS is open at the bottom is its use of scheduled commercial airliners to transport mail across the country. Basic transportation of postal items is the lowest level of the "functionality stack" upon which all other USPS services are built. Yet the USPS has been able to change that lowest level with advances in technology. And that change has occurred without requiring a modification in the user interface, which has stayed relatively, if not absolutely, stable.

As these two examples illustrate, openness at the bottom typically reflects influences from the larger environment within which a given system is embedded—in this case from the worlds of technology and economics.

*Open at the top.* When we say that a system of systems is open at the top, we refer to the system's ability to serve as a basis for an unlimited range of services. In the case of the USPS, consider the following three activities that depend on its services: (a) the mail order industry, (b) stamp collecting, and (c) chain letters

Just as web-based companies like Amazon.com exist only because the internet exists, the giant mail order houses such as Sears Roebuck and Montgomery Ward and more modern companies such as Land's End, LL Bean, SkyMall, Sharper Image, Harry and David, etc. grew up as catalog mail order companies. Each of these companies provides a service that rides on top of the service provided by the USPS. (Some even order from each other.) USPS services are essential (a) for delivering catalogs, (b) for receiving orders, and often (c) for delivering ordered products. Just as there are now many internet-only businesses, there were once many catalog-only mail-order businesses, which operated within the environment provided by the USPS.

Stamp collecting and the businesses that support it exist only because (a) the USPS issues stamps and (b) people decided that stamps have a "collectable" value outside their use to pay for the delivery of items. This is an entirely new and creative application of services offered by the USPS. Of course stamp collecting is not limited to USPS-issued stamps. But the idea here is that stamp collecting is an application that was built on top of and was tangential to the reason the USPS issued stamps. Illustrative of how one service builds on another note that many stamp collecting companies are mail-order businesses and that the USPS (and the postal services of many countries) now issue stamps with the express intent of selling them to collectors.

Chain letters also illustrates how new phenomena may be built on top of an existing USPS service. VanArdsdale [9] has written a comprehensive but apparently otherwise unpublished study of (paper) chain letters. It discusses the wide range of motivations that drive chain letters, e.g., profit, luck, etc. It also describes the historical rates at which chain letters have propagated. (They seem to be making a comeback as the number of email chain letters has declined.) Underneath all this, a feature common to all successful chain letters is the reproduction and distribution of the chain letter contents. Thus one of the emergent functions of the USPS is replication and distribution of content. A chain letter is one way to achieve this end.

The flourishing ecology of services built on top of the services offered by the USPS "lives" in the environment created by the USPS—much as an even more flourishing ecology of services "lives" in the environment created by the internet. It is in this sense that the environment provided by the USPS is open at the top and allows new systems to be developed which offer additional services.

*Always evolving—but slowly.* When we say that a system is always evolving, albeit slowly, we refer not only to its continual evolution as an ecology of services (as just discussed) but also to its relative but not absolute stability with respect to its fundamental service interfaces. Zip codes (and then Zip+4 codes) and fixed 2-character state abbreviations were added to the semantics of postal address in the mid-20$^{th}$ century. Because the semantics of addressing has changed relatively slowly (one might say glacially), users of USPS services were able to accommodate.

One may think of the allowable format for addresses as comparable to the standards and protocols in today's internet environment. The USPS addressing standard has been stable enough to allow users to learn how to use it and to adopt it economically. But the standard is not eternal. It has changed as new demands were placed on the system.

## 2.2 Other communication systems

We have focused on the USPS. Similar (and more dramatic) stories can be told about other wide-area communication systems. Not only are the telephone system and the internet much more open at the top and the bottom, these systems too are in a constant state of change. But they too are stable enough to be useful.

These other systems differ from the USPS in that control over them is far more distributed. The USPS is, after all, a monolithic system owned and operated by an agency of the US government.. Control of the telephone system and the internet is much broader. And they are not run by the government.

We chose to focus on the USPS to illustrate that even a traditional, familiar, and low tech system may be seen as a successful systems of systems. We also thought it worth noting that it is possible for the government to nurture and sustain a viable and long lived system of systems. (Would anyone ever have imagined that the USPS would be held up as a positive model in a technical paper?)

## 2.3 The economic system as a system of systems

Consider the laws and rules regulating the economic system. Focus specifically on laws regulating the formation of corporations and laws regulating how corporations and individuals interact with each other as economic entities.

The system framed by those laws and rules is an environment within which there has been (and continues to be) enormous activity and creativity. This environment, along with the activity that occurs within it, may be seen as a system of systems in the sense described above.

- *It is open at the top.* There are few limits to the products and services that can be created and offered.
- *It is open at the bottom.* There is no lowest economic level of abstraction. Technology at any level down to quantum physics can be pressed into economic service.
- *It evolves continually, but slowly enough to be useful.* The laws and regulations that govern economic activities are not fixed. Old laws are revised, new laws are created, and new categories of economic interests are brought under the law. Intellectual property law is in particular ferment these days.

As the overseer of the economic system of systems the government provides relatively minimal services. It provides a court system, and it provides a monetary system. It also provides certain regulatory services. But for the most part, the economic system consists of non-governmental participants interacting with each other—but within the framework of the system's laws and regulations.

The economic system is enormous in terms of the activities that take place within it. In comparison, the government's participation is relatively small. Even considering the court system, the central banking system, and all governmental economic-related bureaucracies, the cost to operate these is barely visible when compared to the economic system itself. This illustrates the fact that it does not require a massive infrastructure for a successful system of systems to flourish.

## 2.4 Systems of systems exist within larger systems of systems

Communication systems exist within the economic system. Much of their reason for being is economic. In saying this we do not mean to diminish the importance of person-to-person non-economic communication. But the point to be made here is that many of the applications that grew up within the various communication systems did so for economic reasons. Amazon.com the internet entity is but one face of Amazon.com the economic entity.

If communication systems exist within an economic system of systems environment, within which environment does the economic system exist? One might want to construct a nested structure of environments, but in the end there is one final environment: reality. Every system of systems environment is (a) open at the bottom to the extent allowed by the laws of nature and (b) open at the top to whatever we can imagine and find a way to implement.

*Open at the bottom.* One can make money in the economic system. But if one is not successful in the economic system, one can operate in the real world and simply steal.

The economic system is not isolated from the environment in which it functions. Burglary and armed robbery may not exist as valid operations given the rules and regulations of the economic system. But they exist in the larger system. No system (or system of systems) can be completely isolated from its environment. The economic system is open at the bottom in ways we would prefer did not exist.

A more charming illustration of how activity in one environment can be influenced by forces from its containing environment is the development of markets (e.g., on eBay) for virtual assets from online games. (See Castronova [10].) Multiplayer games—in which players interact in virtual worlds and accumulate virtual assets that have value in those virtual worlds—have developed into widely used services within the internet environment. eBay, also a service within the internet environment, allows participants to buy and sell assets for real money. Given this combination, along with the fact that users of both services also exist in the larger economic world, the buying and selling of virtual assets for real money was inevitable. Some game companies discourage this sort of activity, preferring to keep their virtual worlds closed. Others, acknowledging the virtual impossibility of closing off their virtual worlds from the larger, real world, tolerate it. Some even encourage it and make it a feature of their game.

*Open at the top.* Human beings are remarkably creative. We enjoy what has become a flood of new products and services. This is one way in which the economic system is open at the top. It is also open at the top in ways we would prefer did not exist. The 9/11 hijackers took advantage of the fact that the economic system offers classes in how to fly airplanes. They also took advantage of the fact that airplanes are in effect flying bombs. The hijackers didn't have to manufacture their own bombs. They didn't have to transport them to the sites of attack. The system provided bombs and bomb transport if one only looked at it the right way. The 9/11 hijackers made very creative use of the economic system. To accomplish their objective required only that they bring a few box cutters as weapons from the real world into the economic world. Their use of the economic system is one which we would have preferred they hadn't thought of. But we can't eliminate creative thinking. The economic system is open at the top even for applications we would prefer did not exist.

## 2.5 Service oriented architectures

Frequently one has an intuition about something before one is able to formulate it. The same thing often happens in social and business contexts: businesses drift in a direction which only later is understood to be important. That, we believe, is the situation with service oriented architectures. They are increasingly popular, but no completely satisfying rationale for them has been formulated. Ort [11] offers as good (and current) a definition of service oriented architecture (SOA) as any.

> A *service-oriented architecture* is an information technology approach or strategy in which applications make use of (perhaps more accurately, rely on) services available in a network such as the World Wide Web

If we extend this definition beyond information technology and consider the examples discussed above it's clear that these systems have all been implementations of service oriented architectures. Even the economic system as a whole implements an SOA. The framework within which businesses offer their services to each other may quite readily be understood as a network. Unlike many business fads, the growing popularity of SOA reflects an increasingly clear intuition about how the world works.

## 2.6 A caution for the government

We indicated earlier that systems of systems evolve continually, although slowly. The reason this works, and the reason that the systems that participate in a systems of systems adapt to its changes is that it is worthwhile for them to do so. When the telephone system went to 10 (really 11) digit dialing, everyone adapted because it was easier and cheaper to learn how to dial 10 (or 11) digits than it was to continue to call an operator for "long distance" calls.

Similarly, imagine what would happen if a new and more powerful language for expressing web pages (e.g., a major extension and modification of HTML) were adopted as a formal recommendation by the Internet Engineering Task Force (IETF), the advisory committee for web standards, and supported by the major web browsers. We suspect that within a surprisingly short time, most web pages would be converted to the new standard—and that the conversion would occur with very little fuss.

Many government systems consist of elements that themselves are owned and operated by and for the government. Imagine how difficult it would be to make a similar change in a government owned and operated system of systems. Each participating system would have to be modified to use the new standard. For the government to make all these changes (or pay contractors to make those changes) would require an enormous amount of coordination. Very likely it would also be quite expensive.

In contrast, if the participating systems were owned and operated by entities that were motivated to participate in the overall environment, one wouldn't have to formalize all that coordination. The owners of the participating sys-tems would see it in their interest to ensure that their systems continued to function under the new standards.

Thus when the government develops a system of systems environment, it is important to ensure that the participating systems are owned and operated by entities that have enough of a stake in their participation that they will adapt as the environment evolves. Otherwise, the entire system of systems environment will drift into obsolescence.

# 3 Conclusions

The term *system of systems* should not be used to refer to larger versions of hierarchically organized traditional systems. It should be reserved for environments within which systems may interact and serve as services for each other. Service oriented architectures are thus a valid approach to building systems of systems.

Systems of systems have the properties that (a) they are open at the top (to ever new applications), (b) they are open at the bottom (to new ways of implementing the lowest level functions), and (c) they evolve continually—albeit slowly enough to make it worthwhile for the participating systems to adopt to the changes.

When building a system of systems it is important to ensure that the participating systems participate because it is in their interest to do so. In understanding why it may or may not be advantageous for a system to participate in a system of systems, one must keep in mind the larger environment within which the system of system and its participating systems is embedded. The larger environment is also important for understanding the ways in which the system of system is open (at both the top and the bottom) and subject to pressures to change. A system of systems perspective requires one to look outward from a system to its environment and not just inward from a system to its components.

Although the term *system of systems* should be used to refer to an ensemble consisting of a framework along with some participating systems, we suggest that the term itself is distracting and somewhat unfortunate. By referring to the ensemble as a whole, the term fails to distinguish between the framework and the participating elements. What is most important when building a system of systems is that the framework enable the addition of new systems in such a way that they can build on services offered by other participating systems. An extended discussion of these and related issues may be found in Abbott [12].